\documentclass[usenatbib]{mn2e}

\newcommand{\hoz}{\rm H_2 ~ 1\!-\!0 ~ S(1)}
\newcommand{\hto}{\rm H_2 ~ 2\!-\!1 ~ S(1)}
\newcommand{\hozs}{\rm 1\!-\!0 ~ S(1)}
\newcommand{\htos}{\rm 2\!-\!1 ~ S(1)}
\newcommand{\kms}{\rm km\;s^{-1}}
\newcommand{\cmv}{\rm cm^{-3}}

\title[Near-Infrared Emission from NGC~6822 Hubble V]
{High Resolution Observations of the Near-Infrared Emission from
NGC~6822 Hubble V}

\author[Lee et al.]
{Sungho Lee$^{1,2}$,
 Soojong Pak$^{1}$\thanks{E-mail: soojong@kasi.re.kr},
 Sang-Gak Lee$^{2}$,
 Christopher J. Davis$^{3}$,
 \newauthor
 Michael J. Kaufman$^{4}$,
 Kenji Mochizuki$^{1}$,
 and
 Daniel T. Jaffe$^{5}$\\
 $^{1}$Korea Astronomy and Space Science Institute,
      61-1 Whaam-dong, Yuseong-gu, Daejeon 305-348, South Korea\\
 $^{2}$Astronomy Program in SEES, Seoul National University,
      Shillim-dong, Kwanak-gu, Seoul 151-742, South Korea\\
 $^{3}$Joint Astronomy Centre, University Park,
      660 North A'ohoku Place, Hilo, HI 96720, USA\\
 $^{4}$Department of Physics, San Jose State University,
      One Washington Square, San Jose, CA 95192, USA\\
 $^{5}$Department of Astronomy, University of Texas at Austin,
      Austin, TX 78712, USA\\
}

\begin{document}

\date{Accepted 2004 ?? ??. Received 2004 ?? ??; in original form
  2005 May 27}

\pagerange{\pageref{firstpage}--\pageref{lastpage}} \pubyear{2004}

\maketitle

\label{firstpage}

\begin{abstract}
We observed Hubble V, the brightest HII region complex in the
dwarf irregular galaxy NGC~6822, at near-infrared (1.8--2.4
$\micron$) wavelengths using the Cooled Grating Spectrometer 4
(CGS4) at the United Kingdom Infra-Red Telescope (UKIRT). The line
emission maps of Hubble V show the typical structure of a
photo-dissociation region (PDR) where an ionized core, traced by
compact He I emission (2.0587 $\micron$) and Br$\gamma$ emission
(2.1661 $\micron$), is surrounded by an outer layer traced by
molecular hydrogen (H$_2$) emission. The measured line ratios of
$\hto$ (2.2477 $\micron$) / $\hozs$ (2.1218 $\micron$) from 0.2 to
0.6 and the un-shifted and un-resolved line profiles suggest that
the H$_2$ emission originates purely from a photo-dissociation
region (PDR). By comparing the H$_2$ results with a PDR model, we
conclude that Hubble V includes dense ($10^{4.5} \cmv$) and warm
PDRs. In this environment, most of the H$_2$ molecules are excited
by far-UV photons (with a field strength of $10^{2-4}$ times that
of the average interstellar field), although collisional processes
de-excite H$_2$ and contribute significantly to the excitation of
the first vibrational level. We expect that Hubble V is in the
early stage of molecular cloud dissolution.
\end{abstract}


\section{Introduction}

 The dwarf irregular galaxy NGC~6822 is a member of the Local
Group. Because of its proximity ($d = 500$ kpc; \citealt{mca83}),
we can resolve its molecular clouds and star forming regions on
parsec scales (1 arcsec $\simeq$ 2.4 pc). The Large Magellanic
Cloud (LMC) and the Small Magellanic Cloud (SMC) are much closer
to us than NGC~6822. However, they reside so close to the Galaxy
that the star forming process may be influenced by the tidal
force associated with the Galactic gravitational field.  On the
contrary, NGC~6822 is far more isolated and star formation
processes are dictated only by the local conditions in NGC~6822
itself.

The optical view of NGC~6822 is dominated by a large, well
defined, central bar and many bright HII regions and OB
associations.  The brightest and largest HII region complexes,
Hubble I, III, V, and X \citep{hub25}, are located at the northern
end of the bar. The oxygen abundance measured in the HII regions
(12 + log(O/H) = 8.23; \citealt*{leq79,pag80,ski89}) is 2 times
smaller than the Galactic value (12 + log(O/H) = 8.52 in Orion;
\citealt{pei77}) and between those in the LMC (12 + log(O/H) =
8.43) and the SMC (12 + log(O/H) = 8.02; see \citealt{duf84} and
the references therein).

\begin{figure*}
\includegraphics{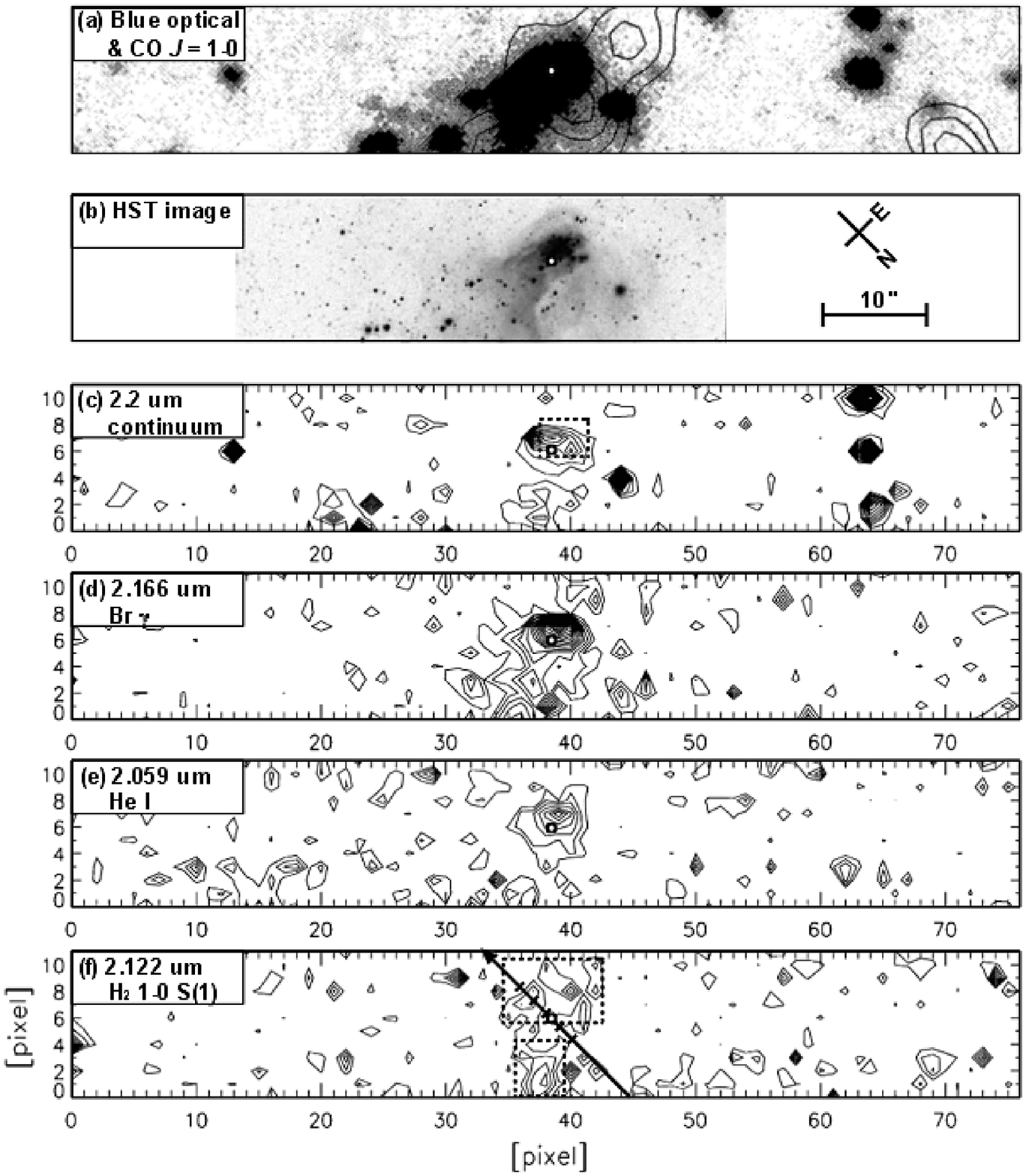} \vspace{193mm} \caption{(a) Blue optical
image and contours of CO $J = 1 - 0$ integrated intensity (from
\citealt{wil94}) over an area covered by our slit scanning
observations with the 40 line mm$^{-1}$ (hereafter l/mm) grating.
The contour intervals are (-4, -3, -2, 2, 3, 4, 5) $\times$ Jy
beam$^{-1}$ (1 $\sigma$). (b) The visual image of NGC~6822 Hubble
V observed by \citet{ode99} and \citet{bia01} using the
NASA/ESA/STScI Hubble Space Telescope (HST). The four other
figures of contours are reconstructed from the scanning of 12
slit positions. The horizontal axis is along the slit and each
row corresponds to each slit. These contour maps show views of
NGC~6822 Hubble V in (c) near-infrared continuum and integrated
line emission of (d) atomic hydrogen, (e) helium, and (f)
molecular hydrogen. Contour levels are linear and increase from
1-$\sigma$ RMS noises with the same intervals; (c) $9 \rm \times
10^{-18}\;W\;m^{-2}\;\micron^{-1}\;arcsec^{-2}$, (d--f) $1 \rm
\times 10^{-19}\;W\;m^{-2}\;arcsec^{-2}$. The open circle at (x,
y) = (38.5, 6.0) pixel marks the position $\rm \alpha = 19^h 44^m
52\fs85, ~ \delta = -14\degr43\arcmin12\farcs8$ (J2000) on the
sky and the pixel scale is 1.22 arcsec $\times$ 1.2 arcsec. The
dotted boxes indicate the areas over which the spectra in
Fig.~\ref{fig_40spectrum} are averaged. The solid arrow in (f)
shows the direction of the slit used for the echelle
observations. The physical conditions at the five positions along
the slit (marked by the ticks; hereafter positions A, B, C, D,
and E from the north to the south) are discussed in the text.
\label{fig_imap}}
\end{figure*}

With a total visible extent of 50 pc (about 20 arcsec on the sky),
Hubble V is the brightest HII region complex in NGC~6822.  Visual
images show the structure of the bright core and the large,
diffuse halo that surrounds the core and extends towards the
northwest (\citealt*{ode99,isr03}; see Fig.~\ref{fig_imap}).  The
core contains a compact cluster of bright blue stars.  The halo is
overlaid on to the eastern part of the OB association, Hodge OB 8
\citep{ode99}. The age of the Hubble V complex is about 4 Myr and
there is no evidence for multiple star formation events in the
past \citep{ode99,bia01}.

\citet{wil94} observed Hubble V in CO emission at high spatial
resolution (6.2 x 11.1 arcsec) and reported finding a molecular
cloud complex in the Hubble V region (Fig.~\ref{fig_imap}(a)).
According to the visual image (Fig.~\ref{fig_imap}(b)) based on a
multi-color composite from the Hubble Space Telescope (HST), it
is likely that dark clouds surround all parts of the core cluster
except in the west. \citet{isr03} found a compact source of
$K$-band emission to the south of the visual core. This $K$-band
peak is not seen in the visual images so they argued that it
could be another compact star cluster that is highly obscured by
the dark cloud.

In this paper, we present the results of our near-infrared
(near-IR) observations of NGC~6822 Hubble V at high spatial
resolutions. The observations and data reduction are described in
Section~\ref{observation}. We present the morphology of the
ionized region and photo-dissociation region (PDR) and the
physical conditions in the PDR in Section~\ref{results}. In
Section~\ref{discussion} we derive some physical parameters by
comparing our observations with predictions of a PDR model
\citep{ste89}. We also compare with previous CO observations to
illustrate the structure of the molecular clouds and discuss the
evolution of the Hubble V star forming region.

\section{Observations and Data Reduction} \label{observation}

We observed the NGC~6822 Hubble V field at the 3.8 m United
Kingdom Infra-Red Telescope (UKIRT) in Hawaii on 2001 June 2--4
and 2004 July 6 (UT), using the Cooled Grating Spectrometer 4
(CGS4; \citealt{mou90}). CGS4 was set up with the 300 mm focal
length camera optics and the long slit of about 90 arcsec. The
observations were obtained at both low and high spectral
resolution: slit scanning at low spectral resolution with the 40
line mm$^{-1}$ (hereafter l/mm) grating was followed by a high
resolution spectrum with the echelle grating. The details are
described in the following subsections.

Initial data reduction steps, involving bias-subtraction and
flat-fielding (using an internal blackbody lamp), were
accomplished by the automated Observatory Reduction and
Acquisition Control ({\sc orac}) pipeline at UKIRT. {\sc iraf}
\footnote{IRAF is distributed by the National Optical Astronomy
Observatories, which are operated by the Association of
Universities for Research in Astronomy, Inc., under cooperative
agreement with the National Science Foundation.} was used for the
remainder of the reduction.  We corrected the spectral distortion
along the dispersion axis using the spectrum of the standard star
BS 7658 as a template. The sky OH lines for the echelle
observation and Argon lines for the 40 l/mm observation were then
used to correct for spatial distortion perpendicular to this axis
and to wavelength calibrate the data.  The telescope was nodded
between on-source positions and carefully-selected sky positions
outside of NGC~6822; the sky frames were subtracted from the
on-source frames to remove OH sky lines. Residual sky lines were
removed by using blank areas on the spectral images; after
subtraction of the sky lines, weak emission of $\hoz$ remained at
measurable levels. At the high spectral resolving power of the
echelle grating the target H$_2$ lines were well separated from
the brightest OH lines so these data do not suffer from this
contamination.  Details pertaining to the low and high-resolution
spectroscopic observations are given below.

\begin{figure*}
\includegraphics{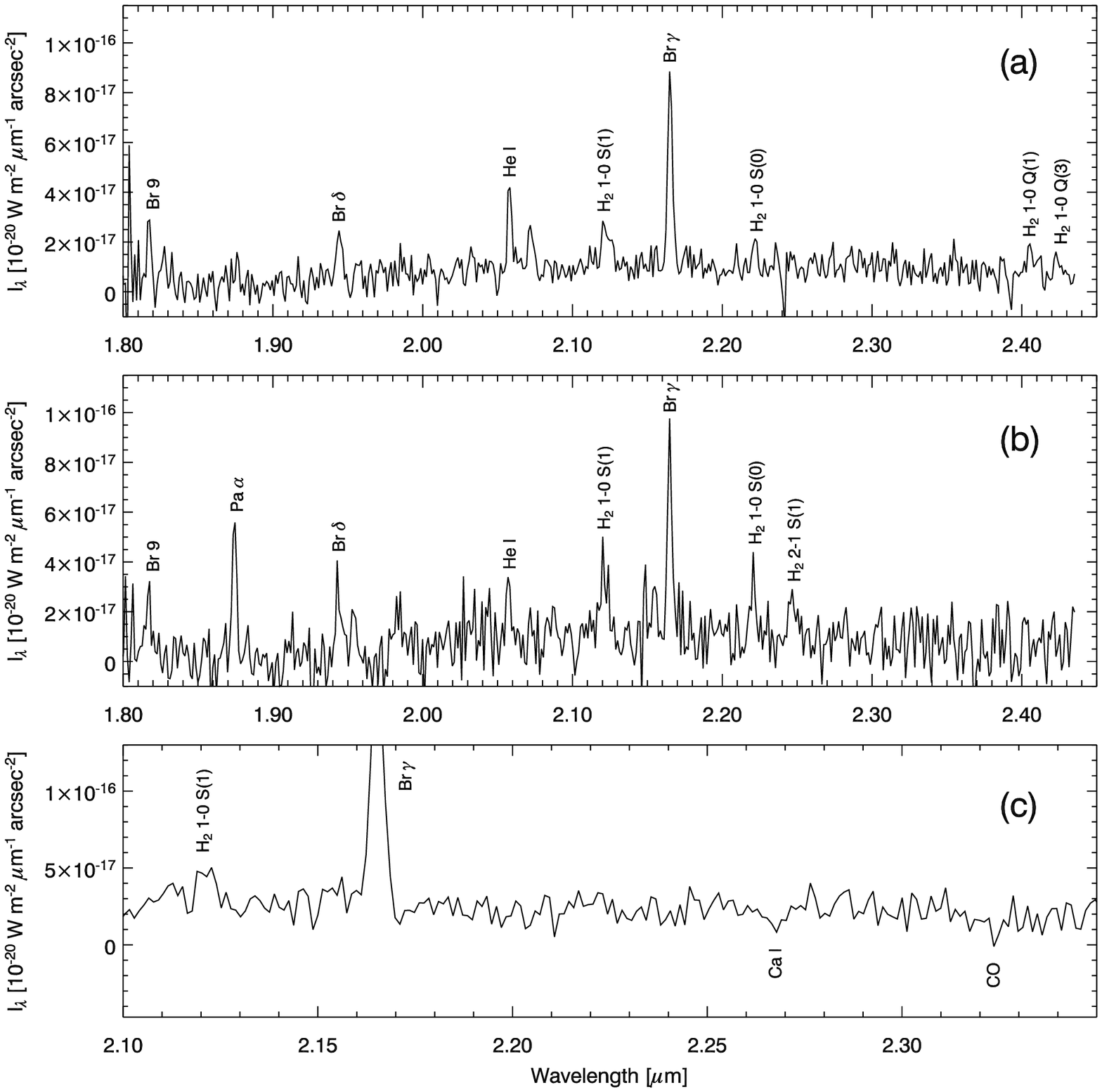}
\vspace{120mm} \caption{$K$-band spectra from the 40 l/mm grating
observations. (a) 40 spectra are averaged over the $9.76 \times
5.41$ arcsec$^2$ area around the core (the upper dotted box in
Fig.~\ref{fig_imap}(f)). (b) 20 spectra are averaged over the
$4.88 \times 5.41$ arcsec$^2$ area in the northwestern halo part
(the lower dotted box in Fig.~\ref{fig_imap}(f)). (c) 12 spectra
are averaged over the visual cluster region of $4.88 \times 3.01$
arcsec (the dotted box in Fig.~\ref{fig_imap}(c)). No correction
for the atmospheric transmission was made. \label{fig_40spectrum}}
\end{figure*}

\subsection{$K$-band spectroscopy : slit scanning} \label{slitscan}

We obtained $K$-band spectra (1.8 - 2.4 $\micron$) with low
spectral resolution ($\lambda / \Delta \lambda = $ 720 - 960)
using a 40 l/mm grating and a one-pixel-wide (0.61 arcsec) slit.
12 parallel slit positions were observed, sampling a $93 \times
14$ arcsec$^2$ area (Fig.~\ref{fig_imap}). The slit was oriented
$45 \degr$ east of north for each measurement; adjacent slit
positions were separated by 1.2 arcsec perpendicular to the slit
length. The pixel size along the slit was 0.61 arcsec and the
seeing was less than 0.44 arcsec. The image quality is degraded,
however, through the optical system of CGS4 and the final spatial
resolution was about 1 arcsec (2.4 pc at the distance of
NGC~6822) according to the FWHM of the flux profile of the
standard star along the slit.

A three dimensional data cube was made by stacking the 12 spectral
images that resulted from the slit scanning. From this cube we
extracted images of the scanned field in individual emission
lines.  Continuum levels were measured on either side (short-ward
and long-ward in wavelength) of each line so that the continuum
emission could be accurately subtracted from each image.
Figs~\ref{fig_imap}(c)--(f) show four contour maps of Hubble V,
in 2.2 $\micron$ continuum, 2.0587 $\micron$ He I, 2.1661
$\micron$ Br$\gamma$, and 2.1218 $\micron$ $\hoz$ emission. Note
that we have binned over two pixels along the slit axis to make
the pixels roughly square in the extracted images. The final
pixel scale in the reconstructed images in Fig.~\ref{fig_imap} is
1.22 arcsec $\times$ 1.2 arcsec.

At low spectral resolution, emission lines from celestial objects
may not be resolved from the nearby telluric absorption lines. In
this case, one cannot make a reliable flux calibration because the
observed emission lines are blended with the telluric absorption
profiles. However, between 2.0 and 2.3 $\micron$, the atmospheric
transmission is nearly 100 percent and most of the emission lines
of interest (2.0587 $\micron$ He I, 2.1218 $\micron$ $\hoz$,
2.1661 $\micron$ Br$\gamma$, 2.2233 $\micron$ $\rm H_2 ~ 1\!-\!0 ~
S(0)$, and 2.2477 $\micron$ $\hto$) are well separated from
telluric absorption lines. On the other hand, at wavelengths
shorter than 2.0 $\micron$ and longer than 2.3 $\micron$,
atmospheric absorption lines are so strong and crowded that we
cannot make use of many important lines, such as 1.8179 $\micron$
Br9, 1.8756 $\micron$ Pa$\alpha$, 1.9451 $\micron$ Br$\delta$, and
the Q-branch of molecular hydrogen around 2.4 $\micron$. All of
these lines can be identified in a spectrum averaged over a large
area (see Fig.~\ref{fig_40spectrum}).

\subsection{Echelle spectroscopy for H$_2$ lines} \label{echelle}

High resolution $\hoz ~ (\lambda = 2.1218 \micron)$ and $\hto ~
(\lambda = 2.2477 \micron)$ lines were obtained using a 31 l/mm
echelle grating and a two-pixel-wide slit centred at $\rm \alpha
= 19^h 44^m 52\fs85, ~ \delta = -14\degr43\arcmin12\farcs8$
(J2000). The slit length was $\sim$~90 arcsec and the orientation
was set to north--south. The position and direction of the slit is
marked as an arrow in Fig.~\ref{fig_imap}(f).

The slit width on the sky was 0.83 arcsec for $\hoz$ with a
grating angle of 64\,\fdg69 and 0.89 arcsec for $\hto$ with an
angle of 62\,\fdg13; the pixel size along the slit was 0.90 and
0.84 arcsec, respectively, for these two configurations. Seeing
was about 0.75 arcsec, but the final spatial resolution after the
CGS4's optics was 2.1 arcsec (5.0 pc) for $\hoz$ and 1.7 arcsec
(4.1 pc) for $\hto$. The instrumental resolutions, measured from
Gaussian fits to the telluric OH lines in our raw data, were $\sim
17~\kms$ for $\hoz$ and $\sim 20~\kms$ for $\hto$, respectively.

\begin{figure}
\includegraphics{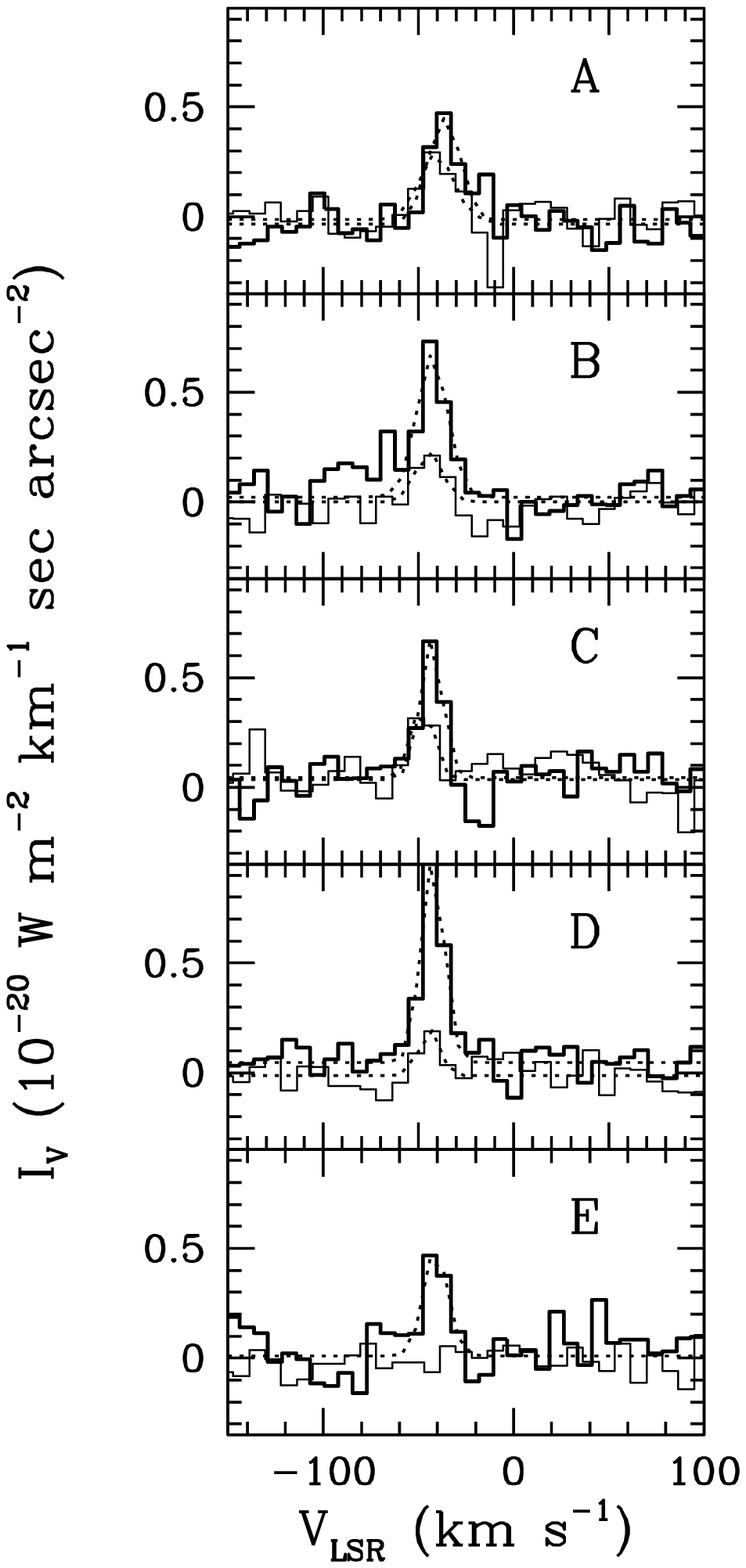} \vspace{105mm}
\caption{
$\hoz$ (thick line) and $\hto$ (thin line) spectra from the high resolution echelle
observations. The position of each spectrum labelled by A--E
is indicated in
Fig.~\ref{fig_imap}(f) and Table~\ref{tbl_fit}. Each spectrum is
averaged over 1.8 arcsec on the sky to improve the signal-to-noise
(S/N) ratios. The dotted lines are Gaussian fits to the observed
line profiles. Note that the spectra are not
corrected for instrumental broadening of $\sim 17~\kms$ for $\hoz$
and $\sim 20~\kms$ for $\hto$, respectively.
\label{fig_spectra}}
\end{figure}

The emitting region along the slit was divided into 5 bins (A--E)
to increase the S/N ratios. The positions of the bins are marked
in Fig.~\ref{fig_imap}(f) and the length of each bin is 1.8 arcsec
(4.3 pc at the distance of NGC~6822). The observed spectra are
presented in Fig.~\ref{fig_spectra}. The emission lines are well
fitted with single component Gaussian profiles and we present the
fitting results in Table~\ref{tbl_fit}. The 1-$\sigma$ errors of
the fitting parameters are estimated based on the RMS noise of the
base line of each spectrum. The average intensity ($\sim$ 1
($\pm$0.1) $\rm \times \; 10^{-19} \; W \; m^{-2}$ $\rm
arcsec^{-2}$; not corrected for interstellar extinction) of our
observed $\hoz$ lines is consistent with the intensity ($\sim$ 0.9
($\pm$0.2) $\rm \times 10^{-19}$ $\rm W \; m^{-2} \;arcsec^{-2}$)
measured by \citet{isr03} using a large, single aperture of 19.6
arcsec in diameter. The intensities and ratios in
Table~\ref{tbl_fit} are corrected for the interstellar extinction.
The adapted foreground reddening $E(B-V)$ to the Hubble V field is
0.65 mag (A$_V = $ 2.02 mag) which is suggested by \citet{isr03}
who compare the radio continuum flux-densities at 1.5, 4.8, and
10.7 GHz to the H$\alpha$ flux.

\begin{table*}
 \centering
 \begin{minipage}{140mm}
  \caption{Gaussian fitting parameters to the H$_2$ line profiles observed by
  the echelle spectroscopy}
  \label{tbl_fit}
  \begin{tabular}{@{}cllllllll@{}}
  \hline
    & & \multicolumn{2}{c}{$\hoz$} & \multicolumn{2}{c}{$\hto$} & \multicolumn{3}{c}{\footnote{All
    the integrated intensities and line ratios are corrected for an interstellar extinction of
    A$_V = $ 2.02 mag.}} \\
   \multicolumn{2}{l}{Position\footnote{
   Relative to $\rm \alpha = 19^h 44^m 52\fs85, ~ \delta =
   -14\degr43\arcmin12\farcs8$ (J2000) along the echelle slit.
   The indications are the same as in Figs~\ref{fig_imap}~\&~\ref{fig_spectra}.}} &
   $v_{LSR}$ & FWHM\footnote{Not corrected for instrumental
   broadening of $\sim 17 \kms$} & $V_{LSR}$ & FWHM\footnote{Not corrected for instrumental
   broadening of $\sim 20 \kms$} & $\rm I_{\hozs}$  &  $\rm I_{\htos}$ & $\htos$
   \\
   &  & \multicolumn{2}{c}{($\kms$)} & \multicolumn{2}{c}{($\kms$)} & \multicolumn{2}{l}{($\rm 10^{-19}\;W\;m^{-2}\;arcsec^{-2}$)}
   &  ~/~$\hozs$  \\
 \hline
 A & N 2.9$\arcsec$ & -36 ($\pm$2) & 20 ($\pm$4) & -41 ($\pm$2) & 20 ($\pm$5) & 1.27 ($\pm$0.30) & 0.78 ($\pm$0.28) & 0.62 ($\pm$0.26) \\
 B & N 1.1$\arcsec$ & -43 ($\pm$1) & 19 ($\pm$3) & -44 ($\pm$3) & 16 ($\pm$6) & 1.60 ($\pm$0.31) & 0.48 ($\pm$0.23) & 0.30 ($\pm$0.15) \\
 C & S 0.7$\arcsec$ & -43 ($\pm$1) & 14 ($\pm$2) & -48 ($\pm$2) & 14 ($\pm$5) & 1.11 ($\pm$0.20) & 0.60 ($\pm$0.28) & 0.54 ($\pm$0.27) \\
 D & S 2.5$\arcsec$ & -43 ($\pm$1) & 14 ($\pm$1) & -44 ($\pm$3) & 13 ($\pm$6) & 1.70 ($\pm$0.22) & 0.35 ($\pm$0.20) & 0.21 ($\pm$0.12) \\
 E & S 4.3$\arcsec$ & -41 ($\pm$2) & 15 ($\pm$4) & --           & --          & 0.94 ($\pm$0.35) & $< 0.29$         & $< 0.31$         \\
 \multicolumn{2}{c}{all\footnote{The spectra of A--E are averaged.}}
                    & -42 ($\pm$1) & 17 ($\pm$1) & -44 ($\pm$1) & 15 ($\pm$2) & 1.28 ($\pm$0.12) & 0.44 ($\pm$0.10) & 0.34 ($\pm$0.09) \\
\hline
\end{tabular}
\end{minipage}
\end{table*}

\section{Results}
  \label{results}

\subsection{Morphology in the Near-IR Bands}
\label{structure1}

The Br$\gamma$ and He I emission maps (see Fig.~\ref{fig_imap})
show the morphology of the ionized region in Hubble V. Around the
core, the emission of these two lines is about the same shape and
size. The Br$\gamma$ emission extends to the bottom of the map,
towards the northwestern halo of Hubble V, where the bright stars
of OB 8 should serve as the source of UV radiation. The halo part
of the diffuse Br$\gamma$ emission is also matched with the He I
emission, although it is very weak. The morphology of our
Br$\gamma$ emission map is consistent with the 1.2822 $\micron$
Pa$\beta$ emission observed by \citet{isr03}. Our slit scanned
Br$\gamma$ image may be more reliable than the Pa$\beta$ image of
\citet{isr03} where the contamination by continuum was not
subtracted completely using a narrow-band filter.

The H$_2$ image in Hubble V (see Fig.~\ref{fig_imap}(f)) shows an
elongated ring (with a size of 18 pc $\times$ 12 pc) that
surrounds the brightest part of the ionized region. This kind of
structure is typical of photodissociation regions (PDRs) in our
Galaxy (\citealt{usu96,ryd98}). The more sensitive echelle
observations (see Table~\ref{tbl_fit}) also show the distribution
of $\hoz$ intensity along the slit which traces a cross-section of
the ring-shaped H$_2$ emission mapped by the 40 l/mm observations
in Fig.~\ref{fig_imap}(f). The intensity distribution has two
peaks, which correspond to the two boundaries of the ring.

The 2.2 $\micron$ continuum image (Fig.~\ref{fig_imap}(c)) traces
the sources in the visual images (Fig.~\ref{fig_imap}(a)\&(b)). In
the extracted continuum image, several field stars evident in the
visual images are clearly seen, while the bright core and the
diffuse halo of Hubble V cover the central part of the map.
However, the compact source at position (x, y) = (64, 1) in
Fig.~\ref{fig_imap}(c) is not identified in the visual images;
instead, this may be related to the nearby CO cloud at (x, y) =
(71, 1), \citet{wil94}'s MC3. The main body of the cloud is likely
more extended than the cloud core traced by CO, since it probably
obscures the compact source.

The distribution of the 2.2 $\micron$ continuum in the core region
is very similar to the $K$-band image observed by \citet{isr03}.
Based on the infrared colors ($J - H = +0.67$ mag and $H - K_s =
+0.16$ mag) derived from their wide band photometry, they
suggested that the $K$-band continuum peak identified by the
visual cluster is dominated by radiaion from $K$ (super)giants. In
$K$-band spectra of late-type (super)giants, the CO bandheads
(e.g. 2.294 $\micron$ $^{12}$CO(2,0) and 2.323 $\micron$
$^{12}$CO(3,1)) should be distinct and seen in absorption
\citep{ram97}. However, we do not detect the $^{12}$CO(2,0)
bandhead and have only a marginal detection of the (3,1) bandhead
in our spectrum of the visual cluster (see
Fig.~\ref{fig_40spectrum}(c)).

A significant fraction of the observed continuum flux can be
explained by free-free or bound-free emission from the ionized
region. In the spectrum averaged over the region around the core
(Fig.~\ref{fig_40spectrum}(a)), the specific intensity at 2.2
$\micron$ is about $1 \rm \times
10^{-17}\;W\;m^{-2}\;\micron^{-1}\;arcsec^{-2}$, while the
integrated intensity of the Br$\gamma$ line is measured to be
$2.94 (\pm 0.17) \rm \times 10^{-19}\;W\;m^{-2}\;arcsec^{-2}$.
These result in the ratio of $I_{\lambda = 2.2
\micron}~/~I_{Br_\gamma} = 34 (\pm2)~\micron^{-1}$ (note that the
scale results from the ratio of a specific intensity and a line
intensity). At $T_e = 11500$\,K \citep{leq79,ski89}, this ratio
is predicted to be about 18 $\micron^{-1}$ \citep{ost89}. Thus,
we can conclude that more than half of the observed continuum
flux is generated by the free-free or bound-free process.

The 2.2 $\micron$ continuum has two peaks in the core. One is
identified with the visual star cluster while the other lies to
the southwest, where there is no visual counterpart in the high
resolution HST image. The position of this southwestern peak is
nearly identical to that of the $K$-band star cluster suggested by
\citet{isr03}. They argued that this new star cluster is brighter
than the visual one but is obscured completely by the thick clouds
of Hubble V. Our 2.2 $\micron$ continuum map shows that the
southwestern peak is indeed brighter than the other peak.

It should be noted that the centres of the ionized regions, seen
in the He I and Br$\gamma$ emission, are not coincident with the
centre of the visual star cluster. Instead, the centres are
probably located midway between the two peaks in the 2.2 $\micron$
continuum map. This implies that the southern, obscured star
cluster is at least as bright as the northern, visual cluster, and
that it is not completely embedded in the dark cloud but emits
strong UV radiation in the direction of the northwestern halo part
of Hubble V. In addition to this, in Section~\ref{model} we will
show that the UV field at the position occupied by the hidden star
cluster is estimated to be stronger than at other positions (see
the derived parameters at position D in Table~\ref{tbl_fit}).

\subsection{H$_2$ Excitation Mechanism} \label{excitation}

\subsubsection{$\hto$ / $\hozs$ line ratio} \label{ratio}

In most cases H$_2$ line emission arises either from thermal
excitation (e.g. by shock heating) or from non-thermal excitation
by far-ultraviolet (hereafter far-UV) absorption
\citep{bla87,bur92,dav00,dav01,pak98,pak04}.
One can in principle distinguish
between these two mechanisms by comparing near-IR line
intensities. The $\hto$ / $\hozs$ ratio has been an effective
discriminant in a number of shocked regions and PDRs. Fluorescent
excitation in a low-density PDR ($n_{H_2} < 5 \times 10^4 \;
\cmv$) should yield a ratio of about 0.6. A lower ratio is
expected in a denser PDR environment, where collisions populate
the levels \citep{bla87,ste89}, or in a shock.

There are two basic types of shock; `jump' or J-type and
`continuous' or C-type (see \citealt{dra93} for a review). J-type
shocks (with velocities greater than about 24 $\kms$) will
completely dissociate the molecules \citep{kwa77}; H$_2$ emission
occurs from a warm, recombination plateau in the post-shock
region.  However, J-type shocks typically produce low line
intensities compared to C-type shocks and $\hto ~ / ~ \hozs$ line
ratios as large as 0.5 are possible because of formation pumping
\citep{hol89}.  At lower shock velocities, below the H$_2$
dissociation speed limit, J-type shocks may yield much lower line
ratios; $<0.3$ \citep{smi95}. In a C-type shock, where the
magnetic field softens the shock front via ion-magnetosonic wave
propagation the H$_2$ dissociation speed limit is much higher
($\sim$ 45 $\kms$; depending on the density and magnetic field
strength in the pre-shock gas). Smaller line ratios of about 0.2
are then predicted \citep{smi95,kau96}. In many astronomical
sources the situation is more complicated, however, and a moderate
ratio may result from a mixture of shocks and PDRs (see e.g.
\citealt*{fer97,lee03,pak04}).

The $\hto$ / $\hozs$ ratios of Hubble V measured from the echelle
observations are presented in Table~\ref{tbl_fit}. It seems that
the ratios at positions A and C are consistent with PDRs. At
positions B, D, and E, however, we cannot distinguish between
dense PDRs, J-type or C-type shocks, or a combination of the two.

\subsubsection{Kinematics of gas motion} \label{kinematics}

Kinematic information can help distinguish between the H$_2$
excitation mechanisms. In a pure PDR environment where the H$_2$
line emission arises from the edges of neutral clouds illuminated
by far-UV photons, the line profiles are narrow. J-type shocks
produce narrow lines and the peak is shifted from the velocity of
the pre-shock gas to that of the shock. C-type shocks, however,
produce broader lines which peak at the velocity of the pre-shock
gas and extend up to the shock velocity.

The H$_2$ spectra observed from Hubble V using the high resolution
echelle grating are presented in Fig.~\ref{fig_spectra}. Observed
line profiles are very narrow (FWHM = 13--20 $\kms$; see
Table~\ref{tbl_fit}). We cannot resolve the lines with the
instrumental resolution of CGS4, which measured 17 and 20 $\kms$
for $\hoz$ and $\hto$, respectively. The measured H$_2$ line
widths are consistent with those of the CO profiles (FWHMs = 4--9
$\kms$) observed by \citet{wil94} and \citet{isr03}. Hence a
C-type shock interpretation may be excluded. On the other hand, we
have found no noticeable shift of the H$_2$ line centres (between
-36 and -48 $\kms$ in $V_{LSR}$; see Table~\ref{tbl_fit} and
Fig.~\ref{fig_spectra}) from those of the CO lines (about -41
$\kms$; \citealt{wil94,isr03}). This result therefore excludes
J-type shocks, because numerous unresolved shocks travelling in
different directions would produce broad H$_2$ profiles. Hence,
the kinematic data, like the excitation analysis, tend to support
a non-thermal excitation mechanism.

\section{Discussion} \label{discussion}

\subsection{Comparing with a PDR model} \label{model}

Given our results that the H$_2$ emission around the core
region of Hubble V arises in a pure PDR environment, we can apply
the observational results to a PDR model to investigate the
physical conditions. \citet{ste89}'s model predicts near-IR
emission spectra of molecular hydrogen in detail for a wide range
of gas density and incident UV field strength. Their model extends
to dense conditions with $n_T \, \ga \, 10^4 \, \cmv$ ($n_T \, =
\, n_H \, + \, n_{H_2}$) where collisional processes affect the
distribution of the H$_2$ ro-vibrational levels, namely
collisional fluorescence, and even farther to the thermal regime.
All results quoted in this section are based on comparisons to
their model.

\subsubsection{Low resolution $K$-band spectrum}
\label{model_40data}

The low resolution spectra from the 40 l/mm observations give us
hints about the physical conditions in the PDR, although
quantitative analyses are difficult due to the low S/N ratios of
the observed H$_2$ lines.

In the spectrum averaged over the region around the core
(Fig.~\ref{fig_40spectrum}(a)), we can identify the H$_2$ lines of
$\hozs$, $\rm 1\!-\!0 ~ S(0)$, $\rm 1\!-\!0 ~ Q(1)$, and $\rm
1\!-\!0 ~ Q(3)$, but cannot detect the $\htos$ line. The
composition of H$_2$ lines like this is consistent with the models
with high density ($n_T \, = \, 10^{\rm 5 \, or \, 6}$ nearly
regardless of $\chi$). The models with low density ($n_T \, = \,
10^{\rm 3 \, or \, 4}$) are inconsistent with the observed
spectrum since the $\rm 1\!-\!0 ~ S(0)$ lines are weaker than the
$\htos$ lines and the total intensities of the Q-branch lines are
lower than those of the $\hozs$ lines in the models. It should be
noted that the Q(1) \& Q(3) lines are suppressed by telluric
absorption and should be stronger than seen in
Fig.~\ref{fig_40spectrum}(a).

On the other hand, in the part closer to the northwestern halo
(Fig.~\ref{fig_40spectrum}(b)), we can observe the $\hozs$, $\rm
1\!-\!0 ~ S(0)$, and $\htos$ lines but cannot detect any Q-branch
line. This spectrum resembles the model spectra of \citet{ste89}
with a low density ($n_T \, = \, 10^3$). The measured line ratio
of $\htos$ / $\hozs$ is $0.5 \pm 0.3$, which is consistent again
with a low-density PDR.

\subsubsection{Echelle data of H$_2$ lines} \label{model_echelle}

Table~\ref{tbl_model} presents the model comparison results with
the observed intensities of the $\hoz$ line and the $\htos$ /
$\hozs$ ratios. In Table~\ref{tbl_fit}, we applied a uniform A$_V
= $ 2.02 mag throughout Hubble V. \citet{isr03}, however, argued
that, to the south of the core star cluster, A$_V$ becomes as high
as 17 mag and more ($E(B-V) \ge 5.4$ mag), because the $K$-band
source cannot be seen in the J and H bands. Hence, we also
compared the data to the models using a different extinction
correction (A$_V = $ 17 mag) at the two southern positions (D \&
E). This, however, does not make any change to our general
conclusion.

The observed intensities are consistent with the models in a range
of $n_T \, = \, 10^{\rm 5 \, or \, 6} \, \cmv$ and $\chi =
10^{2-4}$. Conditions of lower density ($n_T \, = \, 10^{\rm 3 \,
or \, 4} \, \cmv$) cannot produce $\hoz$ intensities as strong as
the observed lines, where $\rm I_{\hozs} \simeq
10^{-19}\;W\;m^{-2}\;arcsec^{-2}$; the models are only marginally
consistent with the observational results if we assume the
strongest UV field ($\chi \, \ga \, 10^4$). However, the $\hoz$
intensity is not expected to increase much more with increasing
$\chi$ in the conditions of low density and high UV field (see
fig.~9 and section~III(c) of \citealt{ste89}), because dust
absorption of far-UV photons dominates over H$_2$ self-shielding
\citep{pak04}. Moreover, the observed $\hoz$ intensities may be
lower limits if the area filling factor is less than unity. Thus,
it should be reasonable to exclude the case of lower density.

\begin{table}
 \centering
 \begin{minipage}{80mm}
  \caption{Gas density and the strength of UV field derived from model comparisons}
  \label{tbl_model}
  \begin{tabular}{@{}clll@{}}
  \hline
   Position\footnote{The indications are the same as in Table~\ref{tbl_fit}.} &
   A$_V$\footnote{Assumed interstellar extinction} &
   $n_T$ ($= \, n_H \, + \, n_{H_2}$)  &  $\chi$ \\
     & (mag) & ($\cmv$)  &  \\
 \hline
 A & 2.02 & $10^{4.3}$ ($10^{4} < n_T \le 10^{4.5}$) & $10^{2-3}$ \\
 B & 2.02 & $10^{4.5}$ ($10^{4.4} \le n_T \le 10^{4.7}$) & $10^{2-3}$ \\
 C & 2.02 & $10^{4.3}$ ($10^{4} < n_T \le 10^{4.6}$) & $10^{2-3}$ \\
 D & 2.02 & $10^{4.6}$ ($10^{4.4} \le n_T \le 10^{4.8}$) & $10^{2-3}$ \\
   & 17   & $10^{4.5}$ ($10^{4.5} \le n_T \le 10^{4.6}$) & $10^{3-4}$ \\
 E & 2.02 & $\ge 10^{4.4}$ & $10^{2-3}$ \\
   & 17   & $\ge 10^{4.5}$ & $10^{2-3}$ \\
 all & 2.02 & $10^{4.5}$ ($10^{4.4} \le n_T \le 10^{4.6}$) & $10^{2-3}$ \\
\hline
\end{tabular}
\end{minipage}
\end{table}

As for gas density, a more precise comparison is possible with the
$\htos$ / $\hozs$ ratio. The ratio is insensitive to $\chi$ and
can be regarded as a function of $n_T$ when $\chi > 10^{2}$ (see
figs~11~\&~12 of \citealt{ste89}). For the case of radiative
fluorescent emission ($n_T \, \simeq \, 10^{3} \, \cmv$), the line
ratios depend on the branching ratios of the radiative cascade. At
$n_T \, \simeq \, 10^{4} \, \cmv$, where the collisional
deexcitation becomes important, and at $n_T \, \ga \, 10^{4.5} \,
\cmv$ with $\chi > 10^{2}$, where the gas becomes warm enough for
collisions to dominate the excitation of the $\upsilon = 1$
levels, $\chi$ does not influence the collisional processes. Our
observational results for the H$_2$ emission point to a
collision-dominated model with $n_T \, \simeq \, 10^{4.5} \, \cmv$
and $\chi = 10^{2-4}$. This means that the region around the core
of Hubble V is equivalent to a dense and warm PDR, where most of
the H$_2$ molecules excited by far-UV photons are collisionally
de-excited and collisional processes contribute significantly to
the excitation of the $\upsilon = 1$ level.

Using the measured radio continuum flux and an assumed HII region
radius of 20 pc, \citet{isr03} found that the molecular gas at the
PDR interface is illuminated by a UV field strength $\chi = 725$.
However, according to the observed distribution of the $\hoz$
emission in this study the front-end of the interface seems to be
much closer to the UV sources, at a distance of about 4 pc.  If
this is the case, the UV flux could be as high as $\chi = 2 \times
10^4$, which is consistent with our expectation at position D
assuming the higher extinction. By comparing the observed CO line
ratios with the PDR model of \citet{kau99}, \citet{isr03} also
suggested a gas density of about $10^{4} \, \cmv$. This is
consistent again with our estimate to within an order of
magnitude.

The \citet{ste89} model used above adopts a metallicity
appropriate for our own Galaxy, whereas our observational results
are for the much lower metallicity environment of NGC~6822.
\citet{kau99} considered a wide range of metallicity in their PDR
model, but they present no prediction for the H$_2$ emission. In
the case of radiative fluorescence with a density of $\la 10^{3}
\, \cmv$, the column density, line intensity, and line ratios of
H$_2$ are known to be nearly insensitive to metallicity
\citep{mal96,pak98}.

\subsection{H$_2$ and CO in the molecular clouds} \label{co}

Our global view of the distribution of the H$_2$ emission is
consistent with the CO map of \citet{wil94}. Both of the H$_2$
features seen in Fig.~\ref{fig_imap}(f), the feature surrounding
the core and the feature extended into the northwestern halo,
could be interpreted as being related to the MC2 cloud of
\citet{wil94}. As for detailed structure, however, this
consistency is not maintained. The CO $J = 1-0$ brightness
decreases from the north to the south by a factor of 2 or 3 (see
fig.~1 of \citealt{wil94}), while the $\hoz$ intensity in the
south is as bright as the emission in the north (see
Table~\ref{tbl_fit}).

The spatial distribution of estimated gas density is shown in
Table~\ref{tbl_model} The gas density seems to increase slightly
from the north to the south. The lower limit at our most southern
position (E) suggests that the gas density does not decrease
towards the south.  This is consistent with \citet{isr03}'s
assumption of high obscuration to the south. However, the $\hoz$
intensity rapidly decreases at the southern-most position. This
contradiction may be explained if the molecular cloud is dense but
has a sharp edge to the south.

The deficiency of the CO $J = 1-0$ emission to the south can be
caused by a geometrical effect. The CO $J = 1-0$ emission is
radiated from surfaces of CO cores since the transition is
optically thick, so the intensity highly depends on the sizes of
the CO cores. In a smaller molecular cloud or under a more intense
UV radiation field where photo-dissociation makes CO cores
smaller, the CO intensity should be fainter. Hence, if we assume
that the southern part of the Hubble V molecular cloud has as high
a density as, but a relatively smaller extent than, the northern
part, and it is illuminated by the obscured star cluster which is
brighter than the visual cluster, then the CO $J = 1-0$ emission
should be significantly reduced at this region.

\subsection{Evolution of the Hubble V complex} \label{evolution}

\citet*{lei89} surveyed 34 Galactic open clusters searching for CO
clouds around each of the clusters. They found that younger
clusters are associated with larger number of more massive and
bigger molecular clouds and that the clouds are receding from
each of the young clusters. \citet{lei91} suggested that the
molecular clouds should be destroyed by their interaction with
newborn, hot stars and further star formation be prevented in the
cluster; massive O stars can rapidly dissolve the associated
clouds by eroding the cloud surfaces with their intense radiation
and by accelerating the clouds systematically with their stellar
winds or via a ``rocket effect'' from evaporated gases
\citep{oor55}.

We can surmise the evolutionary stage of the Hubble V complex by
applying its properties to \citet{lei89}'s empirical relation on
the cloud evolution. \citet{wil94} estimated the virial masses of
the Hubble V molecular clouds, MC1 and MC2, to be $< 4.6 \times
10^4 M_{\sun}$ and $< 6.3 \times 10^4 M_{\sun}$, respectively. If
we adopt \citet{wil94}'s estimations, the Hubble V clouds seem to
be consistent with the Galactic clouds with similar ages of about
4 Myr on the evolutionary relation (see figs 43 \& 47 of
\citealt{lei89}), both in the mass and in the proximity to the
cluster; the Hubble V cluster is overlaid by the molecular clouds
along the line of sight.

At first glance, the sizes of the Hubble V clouds ($<$ 18 pc and
52 pc for MC1 and MC2, respectively, along the major axes),
reported by \citet{wil94}, also look consistent with
\citet{lei89}'s relation. However, the Hubble V clouds should in
fact be significantly larger than the Galactic CO clouds with
similar ages, since a cloud size is defined by the FWHM of the CO
intensity profile in \citet{wil94}, while \citet{lei89} defined
the size using the lowest contour, which is much larger than the
FWHM size. Considering the low metallicity of Hubble V (with 2
times smaller [O/H] than the Galactic value), it seems consistent
with \citet{pak98}'s prediction that the typical size of the star
forming clouds increases as the metallicity decreases.

\section{Conclusions}

We performed near-IR spectroscopic observations of the Hubble V
complex, the brightest HII region complex in the dwarf irregular
galaxy NGC~6822.

From low spectral resolution ($\lambda / \Delta \lambda \simeq
800$) $K$-band (1.8 - 2.4 $\micron$) slit scanning, we obtained
high spatial resolution ($\sim$ 1 arcsec) maps of the region in
2.2 $\micron$ continuum, 2.0587 $\micron$ He I, 2.1661 $\micron$
Br$\gamma$, and 2.1218 $\micron$ $\hoz$ emission. The morphology
of Hubble V in the near-IR is typical of HII regions/PDRs; the
compact He I and Br$\gamma$ emitting region is surrounded by the
H$_2$ emitting region. The detailed distribution of the H$_2$
emission in Hubble V is observed for the first time.

Our high spectral resolution observations ($\lambda / \Delta
\lambda \simeq 15000$) of the $\hoz$ and $\htos$ lines, when
combined with our excitation analysis, indicate that the region
around the core of Hubble V is a dense PDR and suggest that there
is no significant shock activity. The moderate $\htos$ / $\hozs$
ratios (0.2 -- 0.6) are explained by high densities ($n_{H_2} \ga
5 \times 10^4 \; \cmv$) and the possibility of shocks are excluded
by the gas kinematics. The H$_2$ lines have the same systematic
velocity as the cold molecular gas (traced by CO) and the line
profiles are spectrally unresolved. This conclusion implies that
there is no detectable YSO activity around the core of Hubble V.

By comparing the observed results with a PDR model, we estimate a
gas density of $n_T \, \simeq \, 10^{4.5} \, \cmv$ and incident
UV field strength of $\chi = 10^{2-4}$. This means that the
environment around the core of Hubble V is dense and warm enough
so that most of the H$_2$ molecules excited by far-UV photons are
collisionally de-excited and the excitation to the $\upsilon = 1$
level is dominated by collisions. The physical parameters
estimated in this work are consistent with those independently
derived by \citet{isr03}; the strength of the UV field estimated
from radio continuum flux densities and the gas density estimated
from CO line ratios assuming a PDR environment.

The distribution of the near-IR continuum and the line emission
from the ionized regions in the southern part of Hubble V confirm
the existence of a hidden star cluster inside a dark molecular
cloud. Hubble V seems to be in the early stage of molecular cloud
dissolution after having finished its star formation activity
$\sim 4$ Myr ago.  But the progress of evolution seems slower than
in the Galaxy, due to the intrinsically larger sizes and masses of
the molecular clouds.

\section*{Acknowledgments}

We thank Thor Wold, Watson Varricattu, and all the related UKIRT
staffs for their excellent supports for our successful
observations. We also deeply appreciate the efforts of Marc
Seigar and Paul Hirst for the service observations which make our
survey complete. Kind answers from Andy Adamson, Paul Hirst, and
Tom Kerr about the CGS4 were critically helpful for this work. We
are also indebted to Eon-Chang Sung, Yong-Sun Park, and Dae-Hee
Lee for their valuable discussions. Fig.~\ref{fig_imap} is
reproduced from fig.~5 of \citet{wil94} by generous permissions of
Christine D. Wilson and the AAS. SL is especially grateful to CJD
and Tae-Soo Pyo for their warmest care during his visit to Mauna
Kea. The United Kingdom Infrared Telescope is operated by the
Joint Astronomy Centre on behalf of the U.K. Particle Physics and
Astronomy Council.

\bsp

\label{lastpage}

\end{document}